\documentclass[aps,prl,twocolumn,showpacs,superscriptaddress]{revtex4-1}

\usepackage{bbm}
\usepackage{amsmath}
\usepackage{hyperref}
\usepackage{graphicx}
\usepackage{graphics}
\usepackage{amssymb}
\usepackage{mathtools}
\usepackage{xcolor}
\usepackage{physics}
\usepackage[british]{babel}
\usepackage{csquotes}
\usepackage{graphicx}

\graphicspath{{Figs/}}

\begin{document}

\title{Deterministic generation of entangled photonic cluster states from quantum dot molecules}

\author{Arian Vezvaee} 
\affiliation{Department of Physics, Virginia Tech, Blacksburg, Virginia 24061, USA}
\affiliation{Department of Chemistry, University of Colorado Boulder, Boulder, CO 80309, USA}

\author{Paul Hilaire}
\affiliation{Department of Physics, Virginia Tech, Blacksburg, Virginia 24061, USA}

\author{Matthew F. Doty}
\affiliation{Department of Materials Science and Engineering, University of Delaware, Newark, Delaware 19716, USA
}

\author{Sophia E. Economou}
\affiliation{Department of Physics, Virginia Tech, Blacksburg, Virginia 24061, USA}


\begin{abstract}
Successful generation of photonic cluster states is the key step in realization of measurement-based quantum computation and quantum network protocols. Several proposals for the generation of such entangled states from different solid state emitters have been put forward. Each of these protocols come with their own challenges in terms of both conception and implementation. In this work we propose deterministic generation of these photonic cluster states from a spin-photon interface based on a hole spin qubit hosted in a quantum dot molecule. Our protocol  resolves many of the difficulties of existing proposals and paves the way for an experimentally feasible realization of highly entangled multi-qubit photonic states with a high production rate.
\end{abstract}


\maketitle

Multi-qubit entangled photonic graph states and cluster states~\cite{Briegel2001} are integral to several applications of quantum technologies. Some of these applications include measurement-based quantum computing~\cite{Raussendorf2001}, quantum communication in quantum networks~\cite{Kimble2008Nature,Azuma2015,Azuma2017PRA}, and quantum error correction~\cite{Schlingemann2001PRA,Bell2014nature}. As flying qubits with robust coherence properties, photons are of particular interest for the creation of cluster states. However, the fact that photons do not interact with one another, which leads to the absence of dephasing that makes photonic qubits attractive, also creates a key challenge for directly generating photon entanglement. Thus, the entanglement between photons should be mediated through an auxiliary system, typically a matter qubit. Solid-state qubits that are optically active are of particular interest, as they can mediate interactions between photons and also be integrated into devices. Various proposals for the generation of entangled photonic graph states from various solid-state quantum emitters have been considered \cite{LindnerPRL2009,Russo2018PRB,Gimeno2019PRL,Zhan2020PRL,Hilaire2021Quantum,Li2022NPJ,Lee2019,Appel2021arxiv,Michaels2021Quantum,EconomouPRL2010}. 

\begin{figure}
\includegraphics[scale=.6]{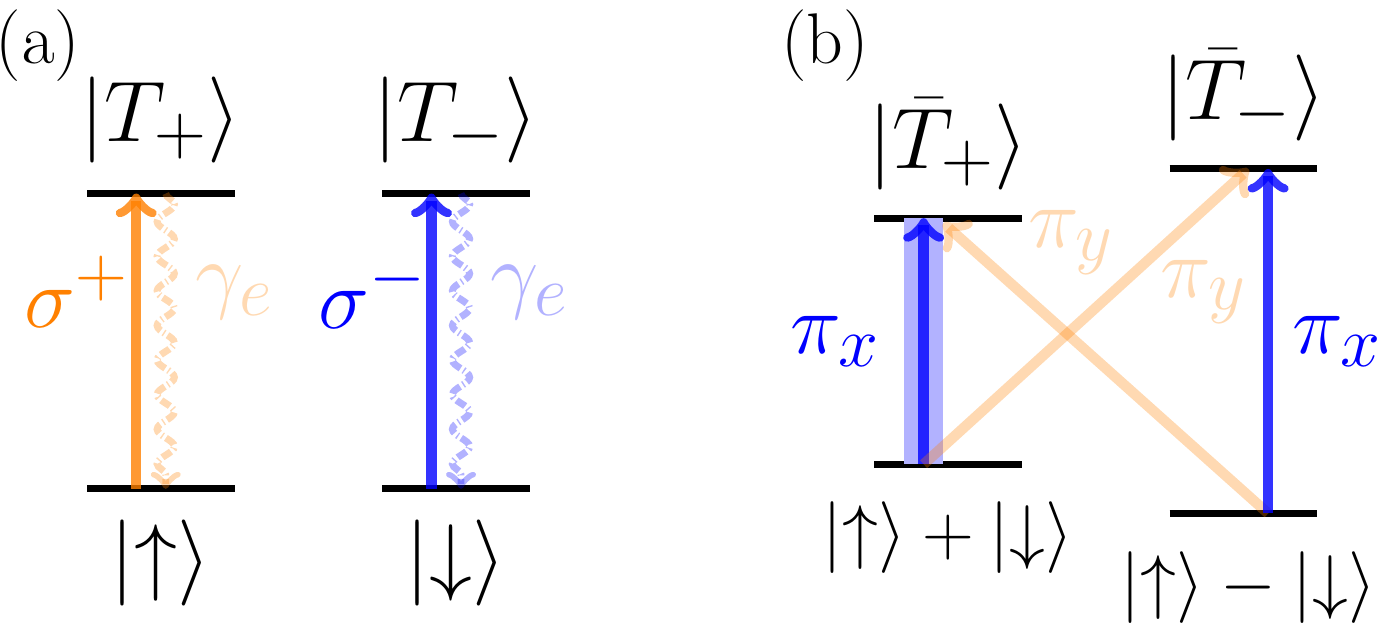}
\caption{(a) The level structure needed for the pumping process of the LR protocol, for the generation of cluster states from a self-assembled quantum dots. Each ground states (electron spins) separately couples to a trion state
using circularly polarized light. (b) The modified selection rules (and corresponding spin and trion states) of a self-assembled quantum dot in transverse magnetic fields. }
\label{fig-LR}
\end{figure}

The Lindner-Rudolph protocol (LR)~\cite{LindnerPRL2009}, is based on a quantum emitter with the selection rules shown in Fig.~\ref{fig-LR}(a). In this scheme the two lower states form the matter qubit, and the system spontaneously emits a photon upon excitation of the qubit states to the excited states. Through the two-step process of alternating between pumping and manipulating the matter qubit, this system will generate a string of photons that are entangled in a one-dimensional cluster state. The LR protocol has inspired several other protocols and experiments. In particular, Schwartz et al.~\cite{Schwartz2016} demonstrated generation of a string of up to five entangled photons using the dark exciton and biexciton states in QDs, which have similar selection rules to those of the LR protocol. This, and similar QD-based experiments~\cite{Lee2019,Vasconcelos2020npj,Cogan2022PRB,Appel2021arxiv}, have faced practical limitations due to the difficulties of perfect realization of the elements of the LR protocol. These challenges include spin dephasing due to the hyperfine interaction with the nuclei, imperfect spin manipulation, and modified selection rules in the presence of transverse magnetic fields. Recently, an alternative experiment, based on a single atom in a cavity, has successfully produced a string of up to 12 photons, by eliminating the mentioned obstacles of solid-state emitters~\cite{thomas2022arxiv}. However, while the `clean' environment of the atomic platforms allow for production of large number of photons, their particularly slow emission rate compared to QDs, could limit the potential applicability of the resulting cluster states ~\cite{Buckley2012reports}.

In this work, we present a cluster-state generation protocol that overcomes the QD-based challenges while taking advantage of their fast photon emission rates. The solid-state emitter is a hole spin qubit in a Quantum Dot Molecule (QDM).  The latter is formed from a pair of vertically stacked self-assembled QDs. This system has been shown to feature a unique level structure, selection rules, and spin properties that are critical for the protocol~\cite{Doty2010PRB}. In particular, the combination of a Lambda system that materializes without the need for an external magnetic field and cycling transitions enables the use of distinct transitions for control and photon emission \cite{Economou2012PRB}. This platform can also give both polarization and time-bin cluster state encodings. Finally, the fact that a hole is used instead of an electron lowers the decoherence that would originate from coupling to nuclear spins in the lattice. As a result, this QDM-based approach not only significantly reduces the experimental overhead, but also leads to much higher-fidelity photonic cluster states. In the following, we  first summarize the limitations of the available protocols. We then proceed to demonstrate how the QDMs will overcome these challenges.

In the LR protocol, devised in QDs (Fig.~\ref{fig-LR}(a)), the two electron spins $\{\ket\uparrow,\ket\downarrow \}$ are defined along the $z$ axis in a magnetic field along the optical axis (Faraday geometry). The two trion states $\ket{T_+}=\ket{\Uparrow\downarrow\uparrow}$ and $\ket{T_-}=\ket{\Downarrow\uparrow\downarrow}$ are constructed from excitonic states that involve combinations of a heavy-hole ($J_z=\pm 3/2$) with two electrons. Angular momentum conservation imposes selection rules using circularly polarized light $\sigma^\pm$. The protocol initializes the qubit state to a superposition state $\ket\uparrow+\ket\downarrow$ and makes use of a periodic train of optical linearly polarized pulses, $\pi_x \sim \sigma^++\sigma^-$, to excite both transitions to a superposition of the two trion states: $\ket{T_+} + \ket{T_-}$. Since the spontaneous emission rate $\gamma_e$ of QDs is fast, the trion superposition state will spontaneously decay to the qubit state by emitting a photon. The emitted photon has equal probability of decaying through both paths, therefore upon emission, the state of photon plus emitter is $\ket{\uparrow}\ket{\sigma^+}+ \ket{\downarrow}\ket{\sigma^-}$ (the pump/emission step of the protocol). Encoding the photon in terms of its polarization and repeating this process $N$ times creates an $N$-qubit GHZ state. 

In order to generate a linear cluster state, an additional control process is required: in between each photon emission we need to apply a $R_Y(\pi/2)$ rotation on the qubit states that leads to a linear cluster state of the photons and emitter (the control step of the protocol). The original LR protocol proposes using an external weak transverse magnetic field to allow the spin to precess in order to implement this $Y$ rotation. Application of such transverse magnetic fields (the so-called Voigt geometry) is of particular interest as it solidifies the coherence time of the electron spin~\cite{Stockill2016Nature}, and allows for an all-optical coherent control of the qubit~\cite{Greilich2009nature,EconomouPRL2007,Economou2006PRB}. However, at the same time, this in-plane magnetic field will cause the precession of the spin projections along the growth axis, which affects the polarization selection rules of the LR protocol, as depicted in Fig.~\ref{fig-LR}(b), where now we have cross transitions in the system. This lowers the fidelity of the resulting cluster state.

To avoid the change of selection rules in order to implement the control step of the LR protocol, in Ref.~\cite{Lee2019} Lee et al. proposed to instead start in a Voigt geometry and selectively enhance one of the transitions to the trion states (highlighted blue transition in Fig.~\ref{fig-LR}(b)) by placing the QD in a cavity and encode the photons in time-bins of emission from this transition. A similar approach was devised by Vasconcelos et al. by driving a single transition of the NV center systems~\cite{Vasconcelos2020npj}. While both of these proposals significantly improve the controllability step of the LR protocol, they are not fully deterministic since the emission step requires a cycling transition. Moreover, in the QD case, there is a non-zero probability of emission from the un-enhanced transition; similarly in the NV center case, there is the possibility of cross-excitation of the other transitions. Furthermore, notice that there is a competition between an efficient photon emission and adequate spin control in the Voigt geometry. Deterministic photon emission requires a strong Purcell enhancement of the desired transition of the $\Lambda$-system compared to the other one. Yet, a strong Purcell enhancement has a detrimental effect for the spin control because it creates an imbalance between the optical coupling of the two transitions. Appel et al.~\cite{Appel2021PRL} have designed a photonic crystal waveguide that induces a cycling transition on one of the vertical transitions of Voigt geometry. However, low-fidelity pulses and far-detuned transition excitation errors have led to low-fidelity states~\cite{Appel2021arxiv}. A full framework considering the potential errors of the time-bin protocol, including the excitation error and imperfect cyclicity, is devised in Ref.~\cite{Tiurev2021PRA}. To summarize, in comparing the time-bin strategy to the original LR protocol we find a competition between emission and control: While the emission of photons works best in a Faraday geometry, the latter is not optimal for the spin control. On the other hand, the Voigt geometry is suitable for spin control, but at the cost of the photon emission.

Another issue to deal with, beyond the controllability and emission, is the dephasing of the matter qubit due to the fluctuations of the nuclear environment. While the efforts to enhance the coherence of  both electron and hole spins in the low-magnetic field limit continue~\cite{Cogan2022PRB}, using the hole spins for the generation of cluster states has the advantage that the dephasing effects will be suppressed: Hole spins have a  weaker hyperfine coupling to the nuclei thanks to their valence band $p$-orbitals. On the other hand, the weak coupling of hole spins to external magnetic fields makes the precession, and consequently the implementation of the rotation, much more difficult. Therefore a direct adaptation of the LR protocol to a hole spin qubit requires strong magnetic fields to implement the rotation on the hole qubits. Additionally, because the trion state has an extra electron it would precess at a much higher rate, which would be detrimental for the LR scheme.

\begin{figure}
\includegraphics[scale=.8]{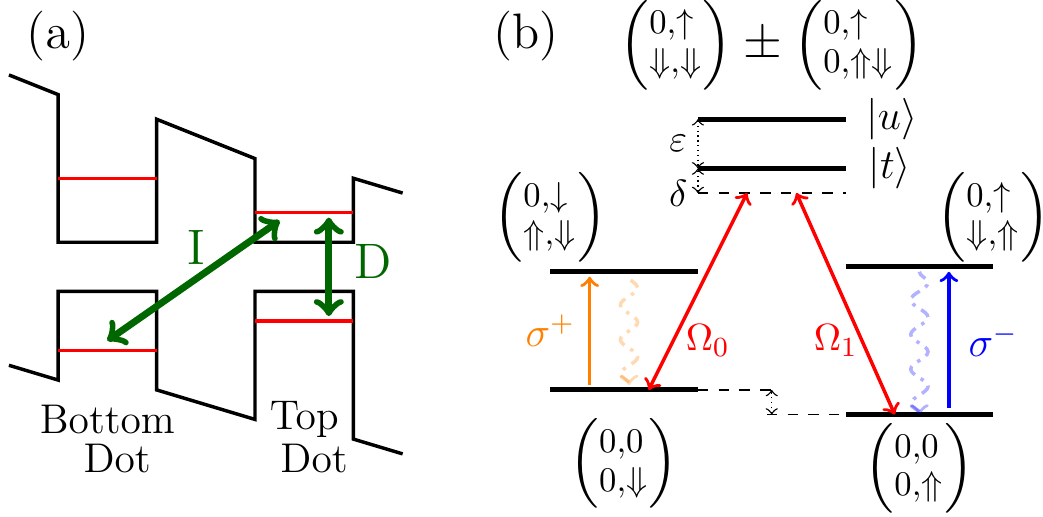}
\caption{(a) Schematic band structure and (b) qubit manipulation strategy for hole-based qubits in QDMs. For perfect generation of cluster states, the $\Lambda$-system in the middle is used for high-fidelity qubit rotations and the two external cycling transitions are used for emission. }
\label{fig-qdms}
\end{figure}

A QDM structure resolves these issues by enabling perfect emission and controllability in a hole spin-based structure that does not require any transverse magnetic fields. QDMs are typically made of InAs and are vertically stacked on top of each other and separated by a barrier, typically made of GaAs. Each dot has individual characteristics (such as their corresponding energy levels),  and the coherent tunneling between them leads to delocalized states that resemble molecules~\cite{Jennings2019qute}. We will use the notation $\tiny{\begin{pmatrix}e_B,e_T\\h_B,h_T\end{pmatrix}}$ to denote the spatial position of particles that reside in each dot. Moreover, this structure enables two different types of transitions: direct transitions that occur within a single dot and indirect transitions that involve charges that reside within two different dots (Fig.~\ref{fig-qdms}(a)). A quantum processor based on this hole qubit is proposed in Ref.~\cite{Economou2012PRB}; it incorporates two important features of QDMs: indirect transitions and hole spin mixing. The qubit states are defined as the two hole spins in the top dot: $\tiny{\begin{pmatrix}0,0\\0,\Uparrow\end{pmatrix}}$ and $\tiny{\begin{pmatrix}0,0\\0,\Downarrow\end{pmatrix}}$. 

Hole states in a single QD receive contributions from all four hole spin projections: heavy holes with $J_z=\pm 3/2$ and light holes with $J_z=\pm 1/2$. However, the ground states are dominated by the heavy hole contribution. In platforms where the two hole spins form the qubit subspace, hole mixing is a necessity for the formation of the $\Lambda$-system that enables coherent control (Fig.~\ref{fig-LR}(b)). A single QD makes use of a transverse magnetic field to achieve a $\Lambda$-system. However, as discussed above, the transverse magnetic field modifies the selection rules and leads to spin precession. On the other hand, in QDMs, the strong spin-orbit interaction combines with a typical slight misalignment of the two dots along the stacking direction to create lead to the intrinsic presence of hole mixing that allows for a $\Lambda$-system \textit{without} a transverse magnetic field~\cite{Doty2010PRB}. Indirect transitions are another important factor for the presented architecture here; they offer significant tunability with local electric fields due to the large static dipole moments caused by placement of charges in two separate dots. This property can be utilized to tune the indirect transitions of several QDMs into resonance with a certain cavity mode.

A control protocol of QDM hole spin qubits, based solely on indirect transitions, is proposed in Fig.~\ref{fig-qdms}(b)~\cite{Economou2012PRB}. The hole spin mixing enables a superposition state with indirect transitions to both qubit states (a $\Lambda$-system), without the need for a transverse magnetic field. This $\Lambda$-system can be used for the implementation of desired rotations between the two qubit basis states. A key additional element of this system is the existence of cycling transitions, which can be used not only for non-destructive spin readout  but, crucially, also to emit spin-entangled single-photons. Notice that these cycling transitions appear naturally in the system, thus, removing the experimental overhead of the need to artificially induce cyclicity in the system through Purcell enhancement. In the following, we demonstrate how we propose to use our novel quantum control methods~\cite{Vezvaee2022arxiv} to allow use of a QDM $\Lambda$-system to implement fast quantum gates with negligible errors.  

The two excited molecular states (labeled target $\ket t = \sin(\eta) \ket{b_1} -\cos(\eta) \ket{b_0}$ and unwanted $\ket u = \cos(\eta) \ket{b_1}+\sin(\eta)\ket{b_0}$, with $\ket{b_1}=\tiny{\begin{pmatrix}0,\uparrow \\\Downarrow,\Downarrow\end{pmatrix}}$ and $ \ket{b_0}=\tiny{\begin{pmatrix}0,\uparrow \\0,\Downarrow\Uparrow\end{pmatrix}}$) are formed due to hole spin mixing~\cite{Economou2012PRB}. Although they have opposite molecular orbitals, both are optically coupled to the qubit states. These levels have similar energies and are separated by the splitting $\varepsilon$. The off-resonant coupling to the unwanted level is quantified in terms of coupling strengths $\lambda_0$ and $\lambda_1$ (the basis state structure relates the two couplings as $\lambda_0 =-\tan(\eta) = -1/\lambda_1$). This unwanted coupling leads to low fidelity of qubit operations. Nevertheless, quantum control methods capable of opposing the phase errors caused by the off-resonant coupling to the unwanted level can be designed. A Coherent Population Trapping (CPT) scheme is used to implement the required $Y$-rotations for the generation of cluster states. By driving the $\Lambda$-system transitions using two identical hyperbolic secant (sech) temporal envelopes that have a $\pi/2$ phase difference, with Rabi frequencies $\Omega_0(t)$ and $\Omega_1(t)$ (denoting the qubit basis sates as $\tiny{\begin{pmatrix}0,0\\0,\Uparrow\end{pmatrix}}\equiv \ket \Uparrow$ and $\tiny{\begin{pmatrix}0,0\\0,\Downarrow\end{pmatrix}} \equiv \ket \Downarrow$), the system can be written in basis consisting of dark state $\ket D=2^{-1/2}(\ket\Uparrow-i\ket\Downarrow)$ and a bright state $\ket B=2^{-1/2}(\ket\Uparrow+i\ket\Downarrow)$, where only the bright state has optically active elements with the excited levels. The optical matrix element defined by the effective Rabi frequency for the target and bright state is $V_{t,B} =\Omega_\text{eff} \sech(\sigma (t - t_g/2))  e^{- i \delta t}$ with gate time $t_g\sim 300$~ps and $\Omega^2_\text{eff} = \Omega^2_0+\Omega^2_1$. For the special case of $\Omega_\text{eff} = \sigma$, this pulse is known to be transitionless~\cite{Economou2006PRB} and it induces a relative phase $\phi= 2 \arctan(\sigma/\delta)$, between the bright and dark states which translates to a rotation about the $Y$ axis in the CPT frame.

For instance, for a detuning below the target level such that $\delta=\sigma$, the CPT framework implements the $R_Y(\pi/2)$. Similarly, we can implement the $R_Y(\pi)$ by setting $\delta=0$. We achieve these ideal evolution operators by modifying the expected detunings to account for the phase errors caused by the presence of unwanted level: $\delta   =  \frac{1}{2}(\varepsilon + \sqrt{\varepsilon^2 + 4 \varepsilon\sigma\cot(\phi/2) -4 \sigma^2})$ to achieve the $R_Y(\phi)$. This modification is inferred by noting that in the case of equal superposition of the two basis states $\eta=\pi/4$, the Hilbert space of the system in the CPT frame transforms into two independent two-level systems, each subject to a transitionless sech pulse. As such we need to modify the detuning to compensate for the phase errors~\cite{Vezvaee2022arxiv}.

\begin{figure}
\includegraphics[scale=.32]{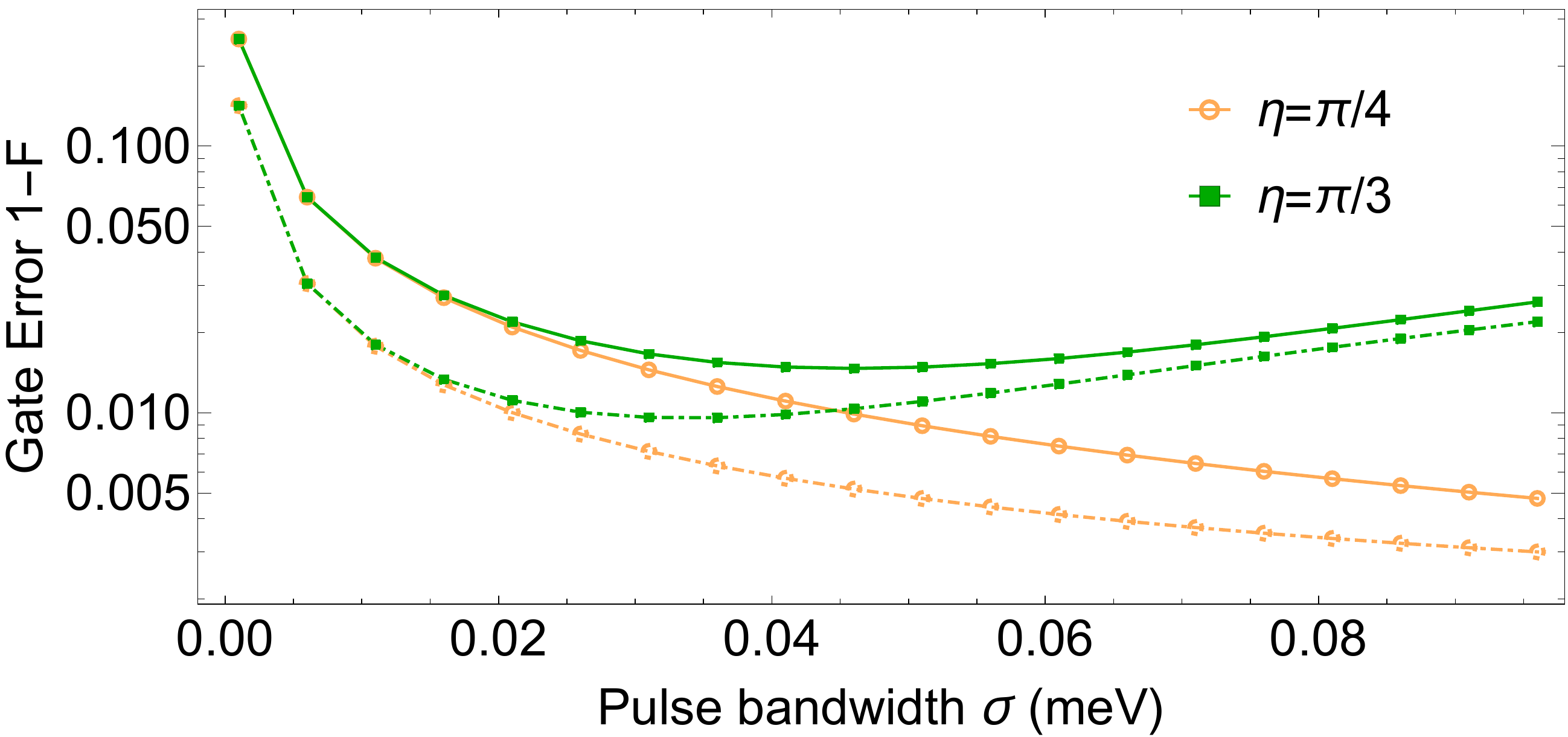}
\caption{The gate error of $R_Y(\pi/2)$ (solid) and $R_Y(\pi)$ (dashed), for different values of couplings to the unwanted level, parameterized by $\eta$. The gates are performed using the $\Lambda$-system in QDMs (Fig.~\ref{fig-qdms}(b)). Unwanted transitions errors are corrected through a simple detuning modification scheme. }  
\label{fig-fidelity}
\end{figure}

Typical values of splitting between the molecular branches in QDMs in which holes tunnel are in the range of ~$\varepsilon=500~\mu$eV and can be controlled with barrier thickness~\cite{Bracker2006APL}. For the pulse to distinguish the two two-level systems, we use a laser bandwidth of $\sigma\sim 0.02$~meV.
Considering typical decay rates of 1 ns (from the direct transition), maximum fidelities of $99.10\%$ for $R_Y(\pi/2)$, and $98.08\%$ for $R_Y(\pi)$ are achievable for $\eta=\pi/4$ (Fig.~\ref{fig-fidelity}). 

The time-bin protocol in QDMs is implemented by using only one of the cycling transitions in Fig.~\ref{fig-qdms}(b). This plays the role of the enhanced vertical transition in a single QD in Fig.~\ref{fig-LR}(b). Notice that this feature of QDMs removes the need for cavity enhancement, and, unlike the single-QD case, has zero probability of photon emission from the competing transition due to the polarization selection rules. Furthermore, the $\Lambda$-system of Fig.~\ref{fig-LR}(b) will be an on-demand source to perform any required rotations without the need for transverse magnetic fields. The protocol for the generation of an $N$-qubit GHZ state, using the ${\sigma^+}$ cycling transition, is as follows:

\begin{itemize}

\item {Step 0:} The system is initialized by creating a superposition state $\ket \psi \sim  \ket \Uparrow + \ket{\Downarrow} $ using a $\pi/2$ rotation through the $\Lambda$-system as described above.
\item {Step 1:} The $\sigma^+$ cycling transition is driven by a $\pi$-pulse, followed by a photon emission to create the first time bin, $\ket \psi \sim{\ket\Uparrow}\ket{0_{\tau=1}}+{\ket\Downarrow}\ket{1_{\tau=1}}$,~where $\ket{0}~(\ket{1})$ denotes absence (presence) of the photon state, and $\tau$ labels the time bins.

\item Step 2: $R_Y(\pi)$ is performed, which results in the state: $\ket \psi \sim\ket\Downarrow\ket{0_{\tau=1}}-\ket\Uparrow\ket{1_{\tau=1}}$.

\item Step 3: The $\sigma^+$ cycling transition is driven again to create the second time-bin entangled state $\ket \psi \sim
\ket\Downarrow\ket{0_{\tau=1}1_{\tau=2}}-\ket\Uparrow\ket{1_{\tau=1}0_{\tau=2}}$.

\item Step 4: Another $R_Y(\pi)$, identical to step 2, leads to: $\ket \psi \sim-\ket\Uparrow\ket{0_{\tau=1}1_{\tau=2}}-\ket\Downarrow\ket{1_{\tau=1}0_{\tau=2}}.$

\item Step 5 (repetition): Steps 1-4 are repeated $N$ times. 

\item Step 6 (encoding): The hole spin is measured in the $\ket\pm\equiv2^{-1/2}(\ket\Uparrow\pm\ket\Downarrow)$ basis.

\end{itemize}

To extend the protocol above to generate an $N$-qubit linear cluster state, in step 4 we implement a Hadamard-like gate ($R_Y(\pi/2)$). Then, during the repetition process (step 5), in the $(N-1)$th round we only repeat steps 1-3 (i.e., step 4 will be dropped). This allows to encode the adjacent time-bins as $\ket{0_m1_{m+1}}\to\ket 0$ and $\ket{1_m0_{m+1}}\to\ket 1$ (for odd $m$). Finally, measuring the hole spin in the $\ket\pm$ basis will lead to an $N$-qubit linear cluster state.
For the LR-like protocol we make use of both cycling transitions. However, note that the two available transitions do not have the same energy and we will need to encode in terms of both polarization and energy. For instance, the state of the qubit and the first emitted photon are $\ket \psi\sim\ket{\Uparrow}\ket{\sigma^+_1,\omega^+_1}+\ket{\Downarrow}\ket{\sigma^-_1,\omega^-_1}$, where the subscripts denote the number of the emitted photon, and the $\omega^\pm$ denote the energy of the photon from the corresponding cycling transition. To achieve purely polarization encoding, and simultaneously achieve more desirable frequencies, downconversion can be used~\cite{DeGreve2012nature}. 



In summary, we have put forward a proposal for the generation of photonic cluster states based on QDMs that uses the unique selection rules of these platforms to overcome the current challenges of available protocols. The presence of natural cycling transitions, distinguished by polarization selection rules, overcomes two of the major sources of errors in time-bin protocol as analyzed in Ref.~\cite{Tiurev2021PRA}: They assure the absence of excitation errors, and allow for unit probability of photon emission from the desired transition (as opposed to $94\%$ probability achieved by means of induced cyclicity~\cite{Appel2021arxiv}). Photon loss remains a critical issue, but advances in site-deterministic QD growth~\cite{McCabe2020JVST, McCabe2021JVST} and photonic device integration~\cite{Borregaard2019AQT, Borregaard2020PRX,Dousse2008,Wang2019, Tomm2021} offer the promise for continued improvement in photon extraction efficiency. QDMs offer the additional advantage of using the wavelength tunability of the indirect (cycling) transitions to further enhance coupling to photonic devices for extraction of the cluster states. Moreover, the capacity of QDMs to implement spin control without transverse magnetic fields is an additional appealing aspect of our proposal. 
Furthermore, our quantum control methods for QDMs allow for pulses that are both fast and have high fidelities. While in the atomic-based emitters, a trade-off between pulse duration (to avoid the cross-talks in the system) and gate fidelity lowers the repetition rate of the protocol, our designed pulses lead to high-fidelity cluster states at orders of magnitude higher rates (ps vs $\mu$s). Moreover, exploiting the fact that  the indirect transitions of multiple QDMs within a single cavity can be tuned into resonance~\cite{Economou2012PRB}, a controlled-$Z$ gate can be implemented between the two QDMs. Using one QDM as an ancila, the generation of multi-dimensional cluster states~\cite{Russo2019NPJ}, and repeater graph states~\cite{Azuma2015} as proposed by Buterakos et al.~\cite{Buterakos2017PRX} can be achieved.

M.D. and S.E.E. acknowledge the support from NSF Grant No. 1839056. P.H. acknowledges support from the EU Horizon 2020 programme (GA 862035 QLUSTER).


\bibliography{biblo}

\end{document}